\documentclass[pre,aps,floatfix,twocolumn,preprintnumbers,superscriptaddress]{revtex4-1}%showpacs
\usepackage{graphicx}% Include figure files
\usepackage{dcolumn}% Align table columns on decimal point
\usepackage{bm}% bold math
\usepackage{amssymb}
\usepackage{pifont}
\usepackage{amsmath}
\usepackage{times}
\usepackage{natbib}
\usepackage{color}
\usepackage{soul}

\def \ii {\text{i}}
\def \dd {{\rm d}}
\def \cc {\text{c.c.}}
\newcommand{\B}[1] {B\bigr[ \theta, \tilde #1(1), \tilde #1(2)\bigr]}
\newcommand{\Bt}[1] {B\bigr[ \theta, \tilde #1(1,t), \tilde #1(2,t)\bigr]}
\DeclareMathOperator{\ee}{\rm e}
\DeclareMathOperator{\acos}{acos}
\DeclareMathOperator{\Real}{Re}

\begin{document}

\title{Noise-induced stabilization of collective dynamics}

\author{Pau Clusella}
\affiliation{Institute for Complex Systems and Mathematical Biology, SUPA,
University of Aberdeen, Aberdeen, UK}
\affiliation{Dipartimento di Fisica, Universit\`a di Firenze, Italy}

\author{Antonio Politi}
\affiliation{Institute for Complex Systems and Mathematical Biology, SUPA,
University of Aberdeen, Aberdeen, UK}

\date{\today}

\begin{abstract}
We illustrate a counter-intuitive effect of an additive stochastic force, which acts independently on each 
element of an ensemble of globally coupled oscillators.
We show that a very small white noise does not only broaden the clusters, wherever they are induced by the 
deterministic forces, but can also stabilize a linearly unstable collective periodic regime: self-consistent 
partial synchrony.
With the help of microscopic simulations we are able to identify two noise-induced bifurcations.
A macroscopic analysis, based on a perturbative solution of the associated nonlinear Fokker-Planck
equation, confirms the numerical studies and allows determining the eigenvalues of the stability problem.
We finally argue about the generality of the phenomenon.
\end{abstract}

% \PhySH

\maketitle

\section{Introduction} 

Typically, noise decreases the {\it coherence} of a dynamical system, by blurring, for instance, a perfect
periodicity, or smoothing the fractal structure of low-dimensional chaos. Furthermore, noise can destabilize an 
attractor, when sufficiently large fluctuations allow overcoming an (effective) energy barrier.
Such effects naturally occur whenever the unavoidable presence of stochastic forces is included into an otherwise 
deterministic evolution. Sometimes, however, noise may unexpectedly have the opposite effect of either increasing
the overall coherence or stabilizing a given dynamical regime.
Well known examples are the stabilization of the inverted pendulum~\cite{Simons-Meerson-2009},
stochastic resonance~\cite{Benzi1981,McDonell2009} and coherence resonance~\cite{Lindner2004321,Pikovsky-Kurths1997}.

In this paper we discuss another such instance, where a finite but small amount of white noise, 
acting independently on an ensemble of identical oscillators, stabilizes self-consistent partial synchrony (SCPS), 
an ubiquitous, collective regime~\cite{Clusella-Politi-Rosenblum2016} observed in ensembles of identical oscillators.
This phenomenon differs from standard noise-induced bifurcations (see, e.g. ~\cite{aumaitre2007noise,luckemoss1989noise}),
where the Lyapunov exponent measuring the local stability of a given regime
changes sign because of the fluctuations induced by a (multiplicative) stochastic force.
Here, the noise acts on the microscopic level, while the regime we are interested in is a collective state.
Macroscopically, SCPS corresponds to a rotation of the probability density of oscillator phases.
Its dynamics, controlled by a nonlinear continuity (Liouville-type) equation, takes place within an 
infinite-dimensional functional space. Depending on the control parameters, 
one or more eigenvalues of the linearized equations (the so-called Floquet exponents) may have a positive real part, 
implying that SCPS is unstable and cannot be thereby maintained indefinitely. 

On the macroscopic level, the effect of noise is described by a diffusion operator, which adds up to the 
continuity equation, transforming it into a nonlinear Fokker-Planck equation. The impact on the overall
dynamics may be relatively trivial, as the regularization of the switching dynamics (see the next section 
for its definition), but also much less so, when it stabilizes SCPS as shown in this paper with the help of
both numerical simulations and semi-analytical calculations.

We mostly focus on an ensemble of Kuramoto-Daido phase oscillators~\cite{Daido-93,*Daido-93a,*Daido-96}, whose 
evolution, in the absence of noise, is entirely controlled by the coupling function $G(\phi)$.
The simplest such example is the Kuramoto-Sakaguchi model~\cite{Sakaguchi-Kuramoto-86}
where $G(\phi)$ is a sinusoidal function. However, in the last years it has been understood
that a purely harmonic coupling is rather special: a few macroscopic variables suffice to describe the 
collective dynamics~\cite{Watanabe-Strogatz-93,*Watanabe-Strogatz-94,Ott-Antonsen-2008}.
At the same time, it has emerged that the addition of a second harmonic suffices to enrich
the resulting phenomenology. For instance, two- (and three-) cluster 
states~\cite{Hansel-Mato-Meunier-93, Kori-Kuramoto-01} have been found in ensembles of identical oscillators,
while a high degree of multistability has been observed in the presence of diversity~\cite{Komarov2014}.
Furthermore, this setup is the minimal one where SCPS can spontaneously 
emerge~\cite{Clusella-Politi-Rosenblum2016}. 

The addition of noise to the bi-harmonic setup has been already studied in Ref.~\cite{vlasov2015synchronization}
for parameter values where, however, no ``shear'' phenomena such as SCPS can arise.
Here, we indeed explore a region where neither the fully synchronous, nor the splay state are stable in the
deterministic limit.
Numerical simulations of the microscopic equations performed for different noise levels show that
a noise amplitude $1.2 \cdot 10^{-4}$ (see the next section for a proper definition) can stabilize SCPS.

For smaller noise, a second regime is present: it is characterized by a pulsating bimodal 
distribution of phases, which can be traced back to the existence of two-cluster states. The transition 
between the two regimes is controlled by a bifurcation that can be either super or subcritical, 
this meaning that there exists a bistability region where, depending on the initial conditions, either 
of the two regimes can be attained.

The details of the bifurcation diagram have been unraveled with the help of a perturbative approach,
the noise amplitude being the smallness parameter. As a result, we have been able to
estimate the deviations induced by the noise on the actual shape of the probability density of phases, 
and then to solve the corresponding linearized equations, to determine the stability properties of SCPS.
These detailed studies of the biharmonic model are accompanied  by a similar (but purely numerical) investigation
of an ensemble of weakly coupled Rayleigh oscillators, where an analogous stabilization of SCPS emerges, 
when a small noise is added to the microscopic evolution equations.

Altogether, the paper is organized as follows. In Sec.~\ref{section:1} 
we introduce the basic model, the biharmonic coupling function,
and recall the phase diagram observed in the purely deterministic limit.
In Sec.~\ref{section:2} we reconstruct the scenario in the presence of noise by performing direct simulations of
the oscillators. There we give pictorial representations of the relevant regimes and confirm the typical signature of
SCPS: a difference between the mean frequency of the distribution and that of the single oscillators.
Sec.~\ref{section:3} is devoted to a thorough perturbative analysis of the macroscopic equations. This includes a general remark on the 
fact that, in the presence of noise, the fully synchronous state becomes conceptually indistinguishable from SCPS.
Finally, a brief analysis of Rayleigh oscillators is discussed in Sec.~\ref{section:4} together with general remarks about future
perspectives.

\section{The model}\label{section:1}

In this section we introduce the main reference model: 
an ensemble of identical, globally-coupled Kuramoto-Daido oscillators,
\begin{equation}\label{eq:system}
 \dot\phi_j=\frac{1}{N}\sum_{m=1}^N G(\phi_m-\phi_j)+\xi_j(t) \; ,
\end{equation}
where $\xi_j(t)$ is a zero-average white noise such that
$\langle \xi_j(t)\xi_m(t')\rangle=2D\delta_{jm}\delta(t-t')$.
The coupling function $G(\phi)$ is assumed to have a biharmonic shape
\begin{equation}\label{eq:G}
 G(\phi)=\sin(\phi+\gamma_1)+a\sin(2\phi+\gamma_2) \; .
\end{equation}
The dynamics of the phase oscillators can be characterized with the help of the order parameters
\begin{equation}\label{eq:order-parameters}
 \tilde P_D(k) =\frac{1}{N} \sum_{j=1}^N \ee^{ik\phi_j}\; . %= \langle \ee^{ik\phi}  \rangle \; .
\end{equation}
The first two parameters  suffice to express the evolution equation (\ref{eq:system}) in 
the more compact form
\begin{equation*}
\dot\phi_j= -\frac{i}{2} \left [ \tilde P_D(1) \ee^{-i(\phi_j-\gamma_1)} + \tilde P_D(2) \ee^{-i(2\phi_j-\gamma_2)} \;
 - \cc \right ] \; .
\end{equation*}
A first important observable used throughout the paper to identify the different regimes is the
microscopic frequency,
\begin{equation*}
 \nu_D= \langle \dot \phi_j\rangle,
\end{equation*}
where $\langle \cdot \rangle$ stands for temporal average, which is the same for all oscillators (no symmetry 
breaking).
Furthermore it is convenient to introduce the mean-field frequency of the ensemble, defined as
\begin{equation*}
\Omega_D=\langle \dot \Phi \rangle\qquad \text{where}\qquad \Phi=\arg\left[\tilde P_D(1)\right] \; .
\end{equation*}

The analysis carried out in Ref.~\cite{Clusella-Politi-Rosenblum2016} has revealed two symmetric parameter regions where,
in the absence of noise,
neither the fully synchronous regime nor the splay state are stable.
Therein, two dynamical regimes have been identified: self-consistent partial synchrony (SCPS)
and a switching dynamics (SD). In the SCPS regime, the probability distribution of oscillator phases 
rotates with a constant velocity without changing shape.
In the SD regime (first discussed in Ref.~\cite{Hansel-Mato-Meunier-93, Kori-Kuramoto-01}), a two-cluster regime is attained,
characterized by oscillations of the cluster widths and of their separation. This regime originates
from a nontrivial form of instability. The behavior of infinitesimal perturbations is controlled 
by the inter-cluster exponent, which measures the response to perturbations of the mutual distance between
the two clusters; and by two intra-cluster exponents, 
which quantify the growth rate of the two cluster-widths. In the biharmonic model the former exponent is negative, 
while the two latter ones have opposite sign. This means that while the width of one cluster increases,
the other decreases exponentially. 
When the width of the wider cluster becomes of order 1, nonlinearities induce an exchange in the order of the
two clusters, so that it starts decreasing. Therefore, the overall stability is controlled 
by the sum of the two intracluster exponents, which is negative in this biharmonic model. 
This means that both cluster-widths oscillate between a value of order 1 and a minimal value, which becomes
progressively smaller: this is nothing but the convergence towards an heteroclinic cycle. When the minimal
width becomes smaller than the computer accuracy, a spurious convergence to the cluster state occurs.
A small amount of disorder among the oscillators (of order $10^{-12}$) eliminates this artificial effect,
giving rise to periodic oscillations which, however, depend weakly on the amount of disorder.

\begin{figure}
\includegraphics[width=0.48\textwidth,clip=true]{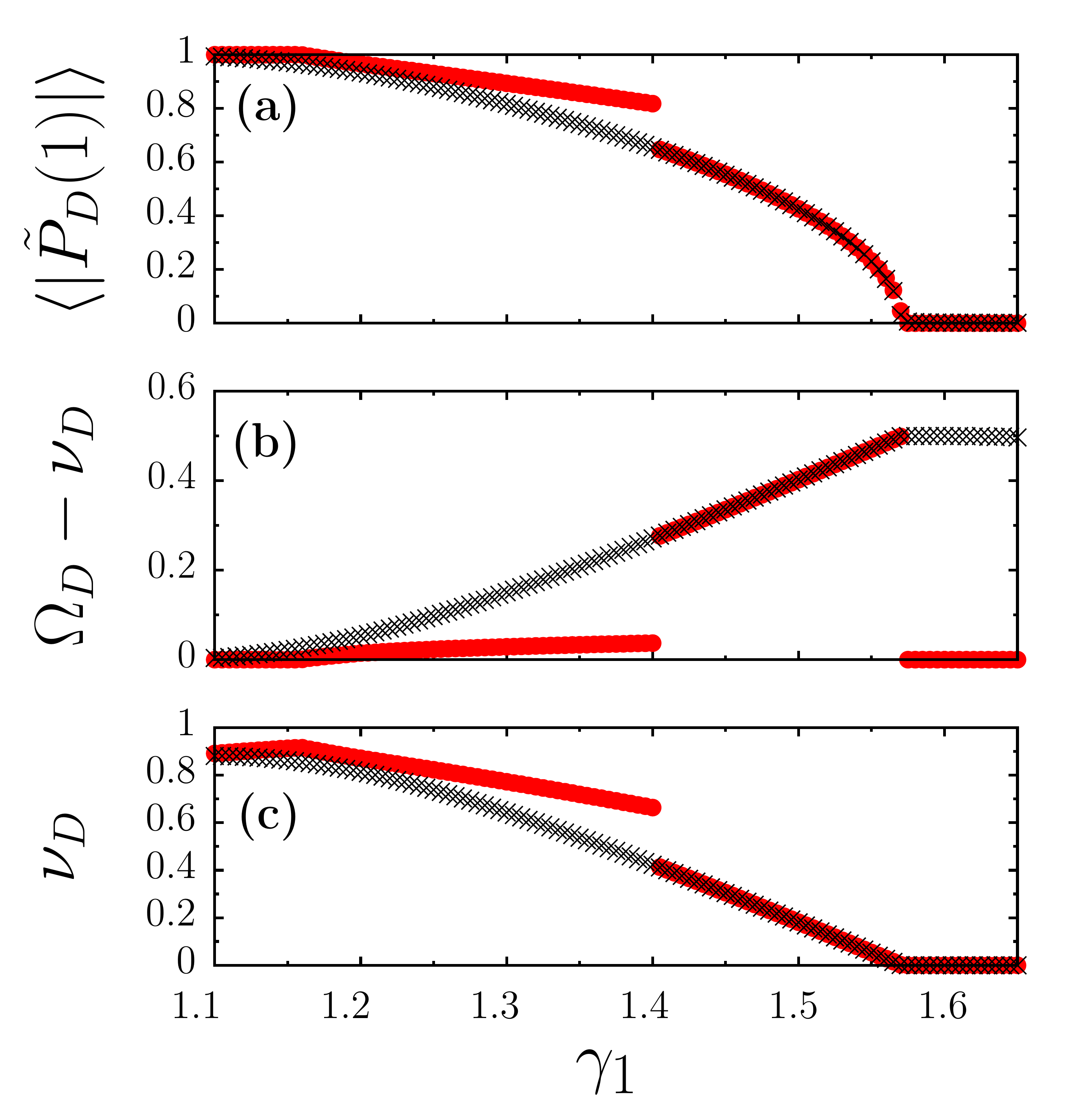}
\caption{Different order parameters versus $\gamma_1$
obtained from direct simulations with no noise at all (red circles)
and $D=1.2\cdot10^{-4}$ (black crosses).
These and all other simulations in this paper refer to $\gamma_2=\pi$.
(a) Time averaged Kuramoto order parameter;
(b) Difference between macroscopic and microscopic frequencies;
(c) Microscopic frequencies of the system, obtained by
averaging over the single frequencies of all the oscillators.
Simulations correspond to the integration of Eq.~(\ref{eq:system}) for $N=1000$ oscillators.
Each order parameter has been computed for $10^5$ time units after discarding a transient of $10^4$ time units.
}
\label{fig:twosimulations}
\end{figure}

For $a=0.2$ and $\gamma_2=\pi$, the parameter regions characterized by these nontrivial regimes are
the intervals $1.159\simeq \acos(0.4) <\gamma_1<\pi/2$ and $3\pi/2<\gamma_1<2\pi-\acos(0.4)\simeq 5.124$.
Here, we will focus on the first one since the second is analogous under the transformations $\phi\to-\phi$ and $\gamma_1\to2\pi-\gamma_1$.
In Fig.~\ref{fig:twosimulations} we show the dependence of various observables on $\gamma_1$ in the
purely deterministic case (see the red circles). In panel (a) we see that upon decreasing
$\gamma_1$, SCPS emerges from the splay state at $\pi/2$
and becomes unstable at $\gamma_1\simeq 1.401$ through a subcritical Hopf bifurcation.
For yet smaller $\gamma_1$-values, a switching dynamics is observed, until the perfectly synchronous state 
becomes stable below $\gamma_1=1.159$.
In panel (b), we can appreciate the typical signature of SCPS: a difference between the microscopic $\nu_D$ and
macroscopic $\Omega_D$ frequency. The SD is also characterized by a frequency difference, although it is so
small it cannot be well appreciated in Fig.~\ref{fig:twosimulations}b.
Finally, the behavior of the microscopic frequency is plotted in panel (c), where one can again recognize 
the loss of stability of SCPS upon decreasing $\gamma_1$.

%%%%%%%%%%%%%%%%%%%%%%%%%%%%%%%%%%%%%%%%%%%%%%%%%%%%%%%%%%%%%%%%%%%%%%%%%
%%%%%%%%%%%%%%%%%%%%%%%%%%%%%%%%%%%%%%%%%%%%%%%%%%%%%%%%%%%%%%%%%%%%%%%%

\section{Microscopic approach}\label{section:2}

\begin{figure}
\includegraphics[width=0.5\textwidth,clip=true]{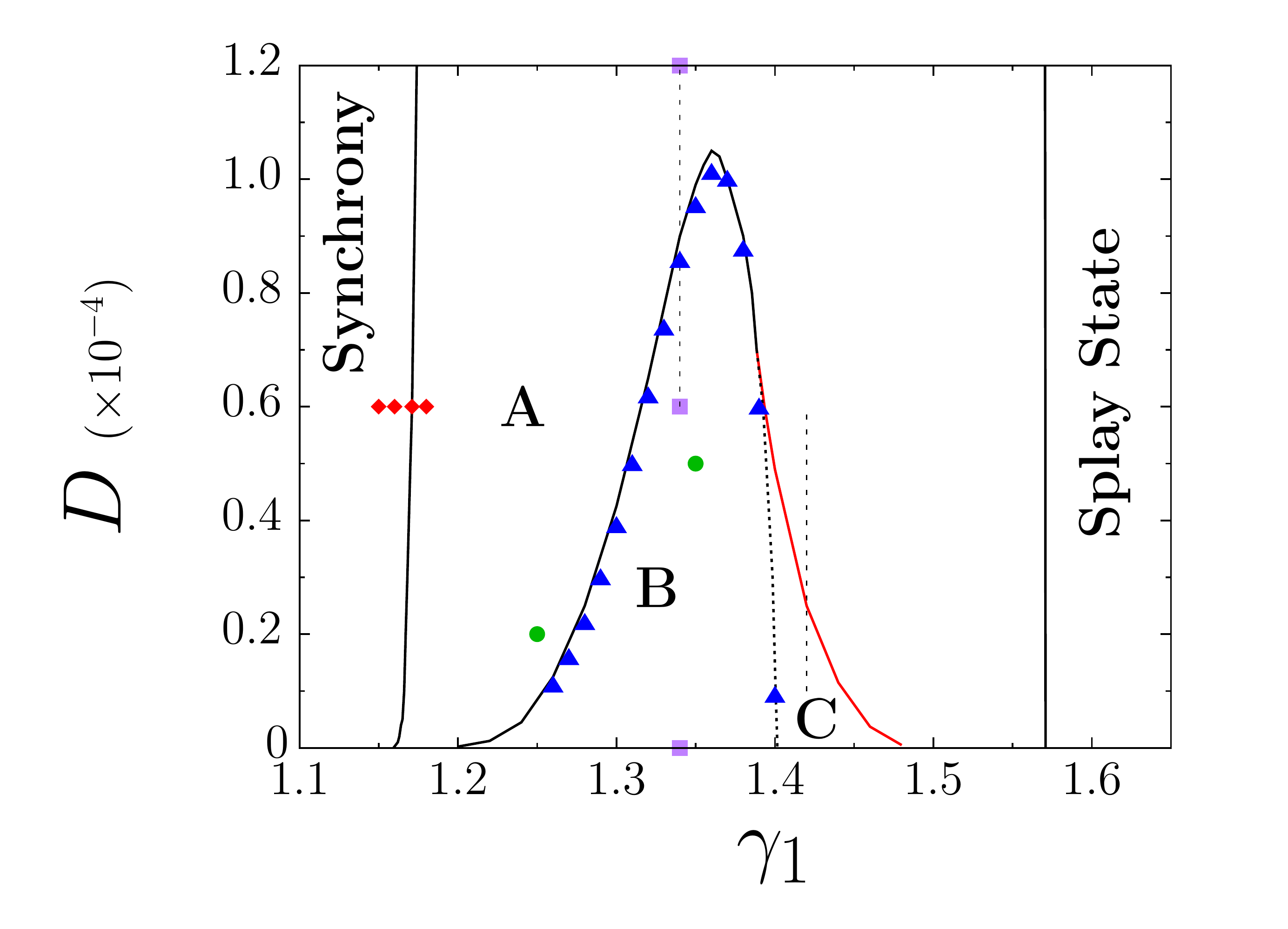}
\caption{Phase diagram in the plane $(\gamma_1,D)$.
In region A only SCPS is stable,
B indicates where only SD is stable and C is a region of bistability.
Black continuous and black dashed lines separating regions A, B and C correspond to super and subcritical Hopf bifurcation respectively, whereas
red continuous line separating regions B and C indicates a saddle-node of periodic orbits bifurcation.
These bifurcation lines have been obtained from direct simulations.
Blue triangles are the results from the macroscopic description developed in section~\ref{section:3}.
The two green circles indicate the points corresponding to simulations in figure~\ref{fig:snapshots}
and the three purple squares correspond to the simulations of figure~\ref{fig:orderparam}.
Vertical black dashed lines indicate the range corresponding to 
figures~\ref{fig:bifurcations} ($\gamma_1=1.34$) and~\ref{fig:bifurcation_jump} ($\gamma_1=1.42$).
The four red diamonds show the parameter values of the potentials in figure~\ref{fig:potentials}.
}
\label{fig:phasediag}
\end{figure}

The very existence of SD reveals that the deterministic dynamics is extremely sensitive to the 
presence of disorder. It is therefore crucial to construct a phase diagram which includes the noise
strength. The results of detailed simulations performed for different values of $\gamma_1$ and $D$
(and different system sizes) are reported in Fig.~\ref{fig:phasediag}.
Five regions can be recognized: on the left synchronous dynamics is observed, while the splay state is
found on the right of the diagram; in between, region A corresponds to stable SCPS, while region B to stable SD;
finally, the two latter regimes coexist within region C.

%%%%%%%%%%%%%%%%%%%%%%%%%%%%%%%%%%%%%%%%%%%%%%%%%%%%%%%%%%%%%%%%%%%%%%%%%%
%%%%%%%%%%%%%%%%%%%%%%%%%%%%%%%%%%%%%%%%%%%%%%%%%%%%%%%%%%%%%%%%%%%%%%%%

On a more quantitative level, we have monitored $\tilde P_D(1)$, $\Omega_D$, and $\nu_D$, 
for $D = 1.2\cdot 10^{-4}$, upon varying $\gamma_1$. 
The results are reported in Fig.~\ref{fig:twosimulations} (see the black crosses), where
the noise is so small that whenever SCPS is deterministically stable,
no appreciable changes are observed.  An important difference with the deterministic case is
that SCPS seems to extend down to the region where full synchrony is stable.
Additionally, we see that the transition from SCPS to full synchrony, expected around $\gamma_1 \approx 1.159$, is smoothed out. 
We anticipate that this is because it is no longer a true bifurcation.
As shown in the next section, the onset of SCPS out of full synchrony can be seen as the tilting of a washboard potential 
beyond the point where a minimum is present. In the presence of noise, an otherwise $\delta$-distribution is broadened,
making it possible to have phase jumps even in the synchronous regime. In other words, the synchronous state is not exactly
synchronous and the microscopic and macroscopic frequencies differ from one another.
Additionally, in Fig.~\ref{fig:twosimulations}(b) we see a curious finite-size effect induced by noise.
In the asynchronous (splay-state) regime, $\tilde P_D(1)$ is not strictly zero for finite $N$. 
One can thereby determine its phase and compute the corresponding growth rate, which, in the presence of noise,
coincides with the frequency $\Omega_D$ of the relaxation oscillations (no such oscillations are generated when
$D=0$).

\begin{figure}
\includegraphics[width=0.45\textwidth,clip=true]{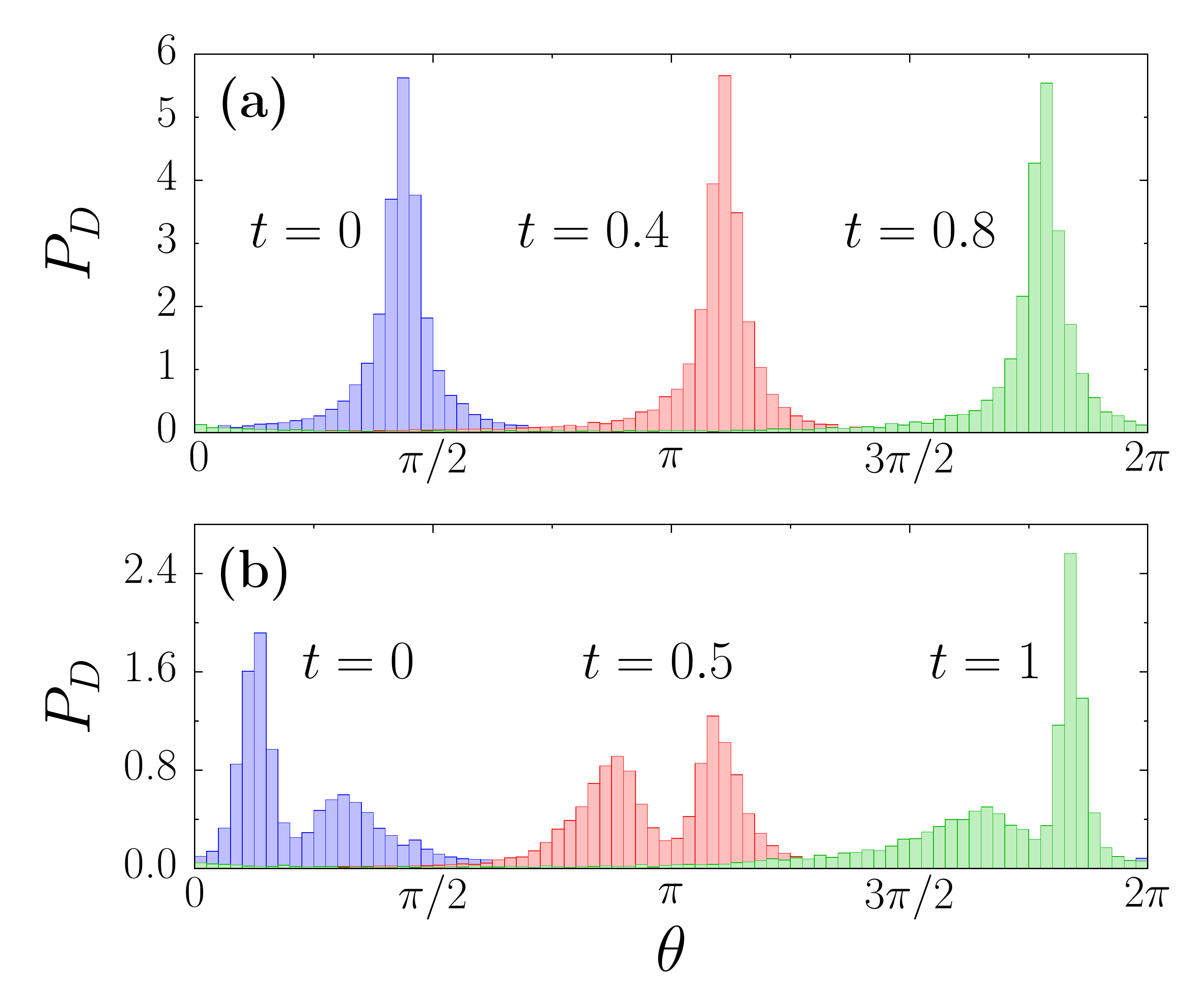}
\caption{Histograms for the density of oscillators at different times.
System parameters are  $\gamma_1=1.25$ $D=2\cdot 10^{-5}$ in panel (a)
and $\gamma_1=1.35$ and $D=5\cdot 10^{-5}$ in panel (b).
The results have been obtained from direct simulations using $10^4$ oscillators after discarding a transient of $10^5$ time units.
}
\label{fig:snapshots}
\end{figure}

In order to illustrate the difference between SCPS and SD, in Fig.~\ref{fig:snapshots} we have plotted a snapshot of the probability
density $P_D(\theta)$ for two different points in parameter space, both falling in the region where SCPS is 
unstable for $D=0$ (see the green circles in Fig.~\ref{fig:phasediag}).
Panel (a) corresponds to the point inside region A: here, the probability density shifts rigidly with a constant
velocity, as expected for SCPS. At variance with the deterministic case, the microscopic quasiperiodicity is obviously lost, 
due to the presence of noise: it still holds true that the microscopic average frequency differs from the macroscopic one. 
Panel (b) corresponds to the green point inside region B: here, the distribution is bimodal
and the two peaks breath -- a reminiscence of the different stability of the two clusters.

%%%%%%%%%%%%%%%%%%%%%%%%%%%%%%%%%%%%%%%%%%%%%%%%%%%%%%%%%%%%%%%%%%%%%%%%%
%%%%%%%%%%%%%%%%%%%%%%%%%%%%%%%%%%%%%%%%%%%%%%%%%%%%%%%%%%%%%%%%%%%%%%%%%%

\begin{figure}
\includegraphics[width=0.45\textwidth,clip=true]{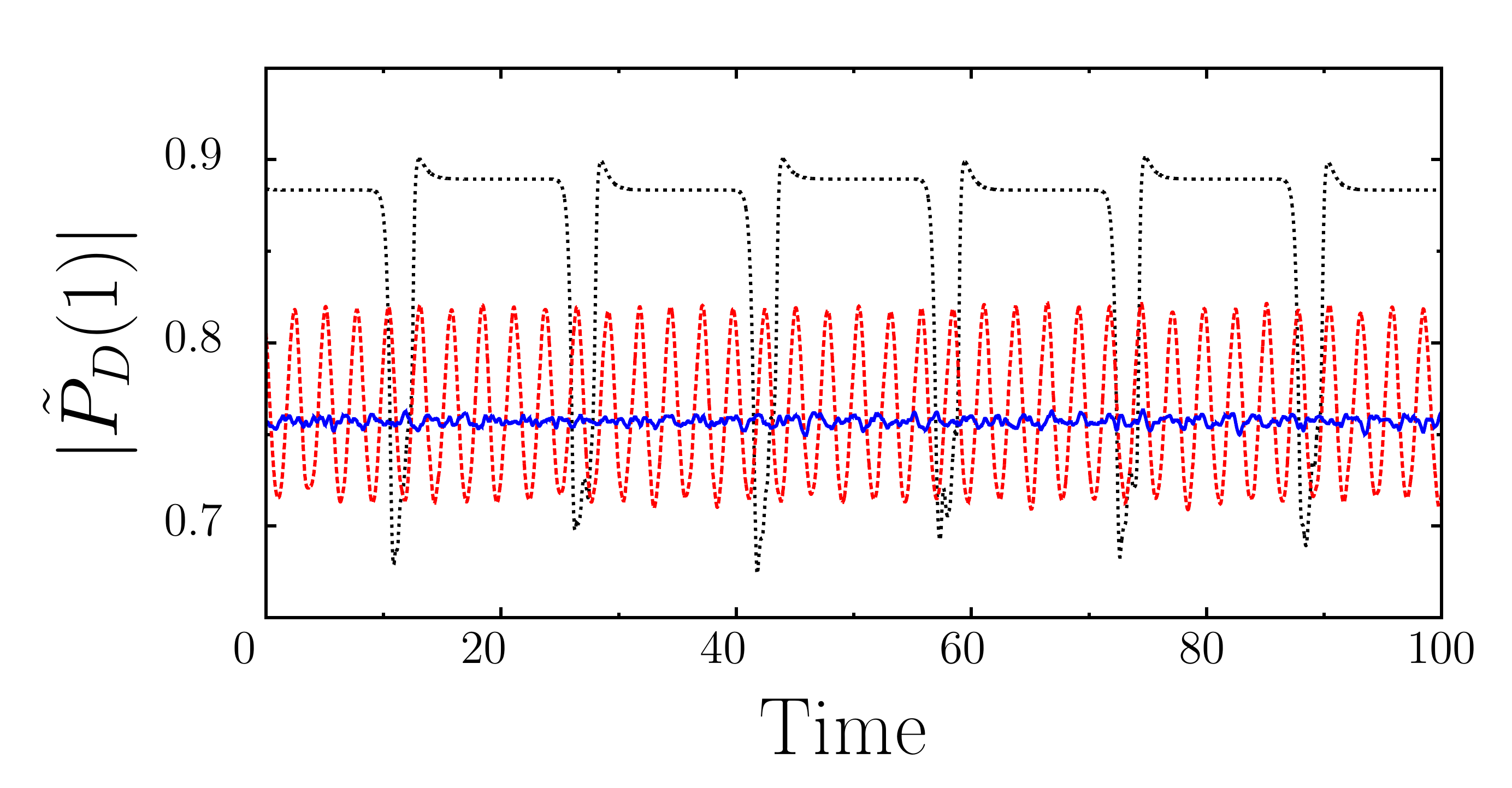}
\caption{Time series of the evolution of the Kuramoto order parameter for
different values of noise and fixed $\gamma_1=1.34$ and $\gamma_2=\pi$.
Black dotted line is the evolution of the deterministic system,
red dashed line corresponds to $D=6\cdot10^{-5}$ and blue continuous line to $D=1.2\cdot10^{-4}$.
Results correspond to simulations with $N=10^4$ oscillators after discarding a transient of length $9\times 10^{4}$.
}
\label{fig:orderparam}
\end{figure}

A more accurate characterization of SCPS and SD is obtained by looking at the time evolution of
the order parameter $|\tilde P_D(1)|$. In Fig.~\ref{fig:orderparam} we present the trace for three points, 
all corresponding to the same $\gamma_1$-value and different noise amplitudes (see the purple squares in Fig.~\ref{fig:phasediag}). 
The black dotted curve corresponds to zero noise (a very small quenched randomness has been added to avoid the spurious collapse 
onto a two-cluster state). Strong, periodic fluctuations are observed, associated to the alternation of contraction
and expansion of the cluster widths. Upon increasing the noise amplitude, periodic oscillations are still observed, the 
amplitude of which decreases while their period shrinks (see the red dashed curve in Fig.~\ref{fig:orderparam}).
A logarithmic reduction of the period was already noticed in~\cite{Hansel-Mato-Meunier-93}: it is due to the
fact that the noise prevents a cluster to become too thin. The additional fluctuations are finite-size effects
which decrease upon increasing the system size. 

For a still small but larger noise ($D=1.2\cdot 10^{-4}$), the oscillations practically disappear: the fluctuations
exhibited by the blue continuous curve are just manifestations of finite-size effects which decrease upon increasing the number of
oscillators. In fact, for this noise amplitude any reminiscence of SD is lost, as confirmed by the series of crosses 
plotted in Fig.~\ref{fig:twosimulations}.

%%%%%%%%%%%%%%%%%%%%%%%%%%%%%%%%%%%%%%%%%%%%%%%%%%%%%%%%%%%%%%%%%%%%%%%%%
%%%%%%%%%%%%%%%%%%%%%%%%%%%%%%%%%%%%%%%%%%%%%%%%%%%%%%%%%%%%%%%%%%%%%%%%

\begin{figure}
\includegraphics[width=0.5\textwidth,clip=true]{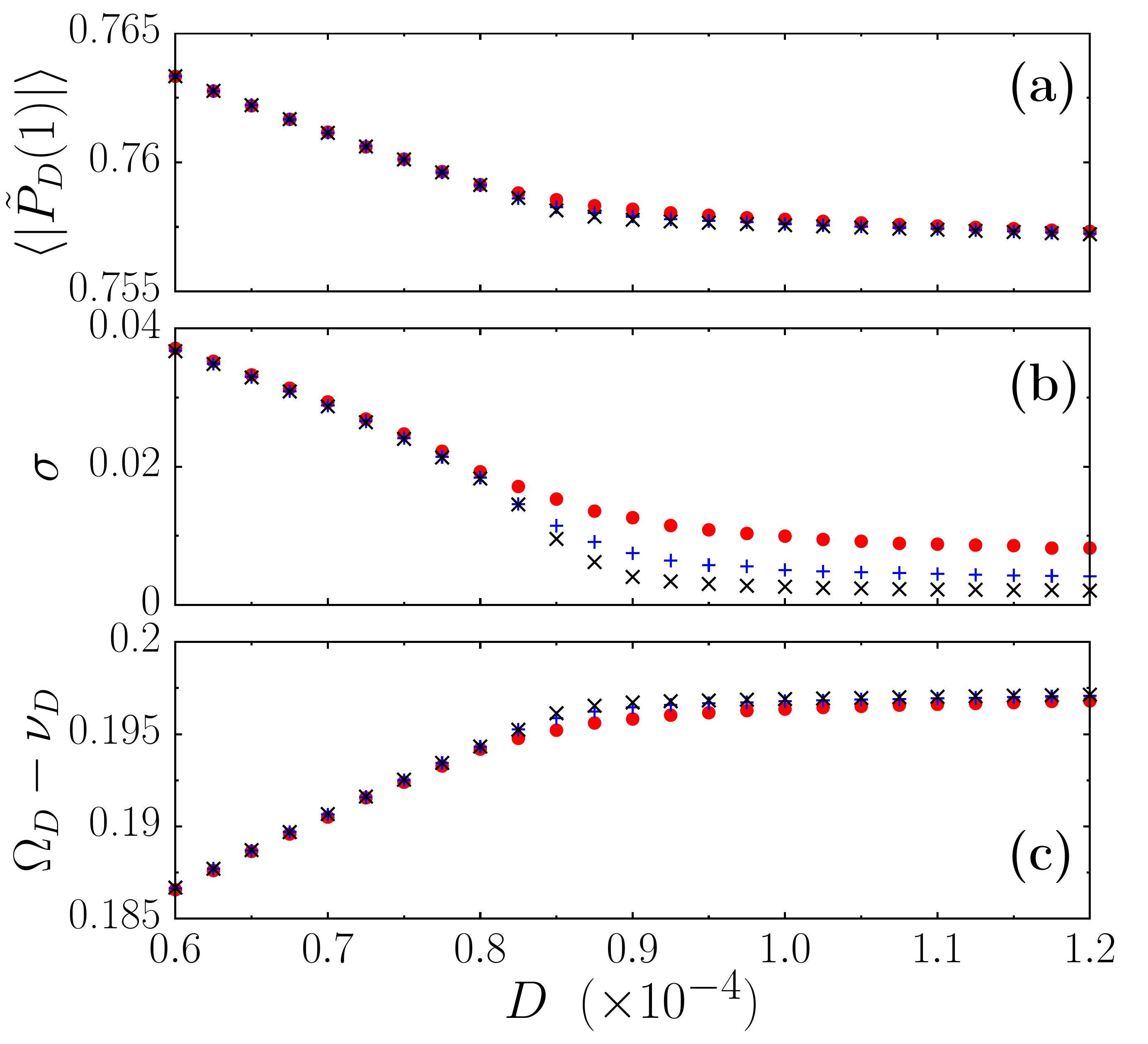}
\caption{Different order parameters extracted from direct simulations of the biharmonic model versus 
the level of noise for $\gamma_1=1.34$.
Panel (a) shows the time-averaged Kuramoto order parameter $\langle \tilde P_D(1) \rangle$;
panel (b) shows its standard deviation $\sigma$; panel (c) shows the difference between the macroscopic frequency 
$\Omega_D$ and the averaged microscopic frequencies $\nu_D$.
Red circles, blue pluses and black crosses correspond to $N=1000$, 4000 and 16000 oscillators respectively.
}
\label{fig:bifurcations}
\end{figure}

Altogether, the direct simulations suggest that the transition from SCPS to SD corresponds to a Hopf bifurcation,
beyond which a constant order parameter starts exhibiting periodic oscillations, which are
reminiscent of the presence of the unstable two-cluster state.
One can interpret SD as a sort of more structured SCPS, since also in this 
case there is a difference between the microscopic and macroscopic frequency. 
In order to validate this interpretation, we have determined the time averaged Kuramoto order parameter $\langle |\tilde P_D(1)|\rangle$ and
the mean-field frequency $\Omega_D$ for $\gamma_1=1.34$ and an adiabatic increase of $D$ from region B to A
(along the left dashed line in Fig.~\ref{fig:phasediag}).
In the top panel of figure~\ref{fig:bifurcations} we see that $\langle |\tilde P_D(1)|\rangle$ progressively decreases upon increasing the noise: 
this is because noise tends to glue together the two clusters. A signature of a true transition can be seen in panel (b)
where the (temporal) standard deviation of $|\tilde P_D(1)|$,
\begin{equation*}%\label{eq:stddev}
 \sigma=\sqrt{\left\langle \left(|\tilde P_D(1)|-\left\langle|\tilde P_D(1)|\right\rangle\right)^2\right\rangle},
\end{equation*}
is plotted for different numbers of oscillators. Upon increasing
$N$, $\sigma$ clearly approaches zero above a critical noise strength.  
Finally, in panel (c) we see that $\Omega_D-\nu_D$ is different from zero both above and below the bifurcation, confirming that
the qualitative difference between SD and SCPS is just the periodic modulation of the Kuramoto order parameter.

%%%%%%%%%%%%%%%%%%%%%%%%%%%%%%%%%%%%%%%%%%%%%%%%%%%%%%%%%%%%%%%%%%%%%%%%%
\begin{figure}
\includegraphics[width=0.47\textwidth,clip=true]{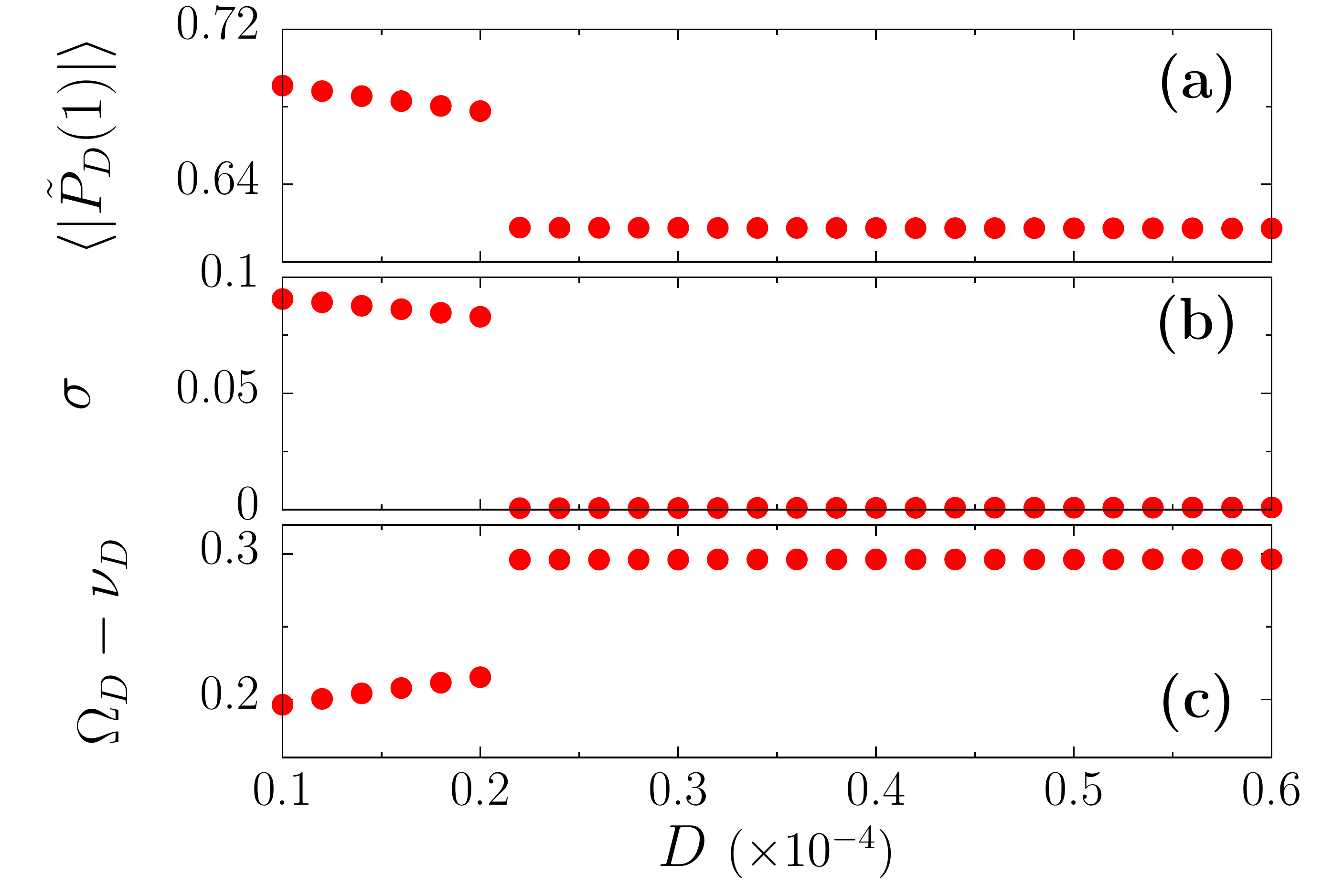}
\caption{Different order parameters extracted from direct simulations of the biharmonic model versus 
level of noise for $\gamma_1=1.42$.
Panel (a) shows the time-averaged Kuramoto order parameter $\langle \tilde P_D(1) \rangle$,
panel (b) shows its standard deviation $\sigma$,
panel (c) shows the difference between the macroscopic frequency $\Omega_D$ and the averaged microscopic frequencies $\nu_D$.
Simulations correspond to $N=10^5$ oscillators,
}
\label{fig:bifurcation_jump}
\end{figure}

%%%%%%%%%%%%%%%%%%%%%%%%%%%%%%%%%%%%%%%%%%%%%%%%%%%%%%%%%%%%%%%%%%%%%%%%%
The transition scenario from SD to SCPS remains essentially unchanged so long as $\gamma_1 \lesssim 1.4$.
For larger $\gamma_1$-values, in region C, both SCPS and SD are stable for arbitrarily small noise
so that one expects the Hopf bifurcation separating B from C to be subcritical.
This scenario is confirmed by the discontinuous jumps observed in 
Fig.~\ref{fig:bifurcation_jump}, where the average order parameter, the standard deviation $\sigma$, and the frequency 
difference are plotted while adiabatically increasing $D$ starting from the SD regime
(along the right vertical dashed line in Fig.~\ref{fig:phasediag}).
In practice SD suddenly disappears through a saddle-node bifurcation, where it collides with a similar unstable regime.

%%%%%%%%%%%%%%%%%%%%%%%%%%%%%%%%%%%%%%%%%%%%%%%%%%%%%%%%%%%%%%%%%%%%%%%%%
%%%%%%%%%%%%%%%%%%%%%%%%%%%%%%%%%%%%%%%%%%%%%%%%%%%%%%%%%%%%%%%%%%%%%%%%%

Upon increasing $D$, these two bifurcation lines bounding region C merge together in a tricritical 
point, a Bautin bifurcation.
The identification of this point requires some extra effort since, close to it, the amplitude of the discontinuity progressively
vanishes. By comparing simulations performed by adiabatically increasing and decreasing $\gamma_1$, we estimate
the tricritical point to be located at $\gamma_1\simeq 1.3885$, $D\simeq 7\times 10^{-5}$.

%%%%%%%%%%%%%%%%%%%%%%%%%%%%%%%%%%%%%%%%%%%%%%%%%%%%%%%%%%%%%%%%%%%%%%%%%%%%%%%%%%%%

\section{Macroscopic description}\label{section:3}

In the previous section we have seen that a small noise stabilizes SCPS with the exception 
of the tiny parameter region B, where periodic oscillations are observed.
Here, we approach the problem from the macroscopic point of view, extending the method introduced in
Ref.~\cite{Clusella-Politi-Rosenblum2016} to account for the presence of microscopic noise.

In the thermodynamic limit, $N\to \infty$, the evolution of the probability density $P_D(\theta,t)$ 
of oscillators with phase $\theta$ at time $t$ is controlled by the nonlinear Fokker-Planck equation
\begin{align}\label{eq:fokker-planck}
 \frac{\partial P_D}{\partial t} &= 
 \frac{\partial }{\partial \theta}\left[\left(\Omega_D-\int\dd\psi G(\psi-\theta)P_D(\psi,t) \right)P_D\right] \nonumber\\
 &+D\frac{\partial ^2 P_D }{\partial \theta^2}\; ,
\end{align}
where $D$ is the diffusion term.
This equation refers to a rotating frame $\theta=\phi-\Omega_D t$. For a properly selected $\Omega_D$-value this equation
admits a stationary solution $P_D(\phi)$, which corresponds to the SCPS regime. It is convenient to introduce the
Fourier representation
 $P_D(\phi)=\frac{1}{2\pi}\sum_{k=-\infty}^\infty \tilde P_D(k,t) \ee^{-ik\phi}$
where 
\begin{equation*}
\tilde P_D(k,t)=\int_0^{2\pi} \dd \psi P_D(\psi,t)\ee^{ik\psi}.
\end{equation*}
In fact, by using this notation, the integral in Eq.~(\ref{eq:fokker-planck}) 
can be simplified, making use of Eq.~(\ref{eq:G}),
\begin{gather}\label{eq:simply}
 \Bt{P_D}:= \int\dd\psi G(\psi-\theta)P_D(\psi,t) \\ 
 =-\frac{i}{2}\left(\tilde P_D(1,t) \ee^{-i(\theta-\gamma_1)}+a \tilde P_D(2,t) \ee^{-i(2\theta-\gamma_2)} -\cc\right) \nonumber
 \; . 
\end{gather}

Notice that the $\tilde P_D(k,t)$ coefficients coincide with the order parameters introduced in Eq.~(\ref{eq:order-parameters}).

A stationary solution of Eq.~(\ref{eq:fokker-planck}) can be obtained by setting the time derivative of $P_D$ equal to zero
and thereby integrating the r.h.s. to obtain
\begin{equation}\label{eq:stationary-equation}
 -H_D=\left\{\Omega_D-\B{P_D}\right\}P_D(\theta)+D\frac{\partial P_D}{\partial \theta}(\theta) 
\end{equation}
where $H_D$ is the probability flux.
The mean frequency $\nu_D$ of the single oscillators 
can be expressed in terms of the flux as $\nu_D=\Omega_D+2\pi H_D$.

Two simple solutions are characterized by a zero flux $H_D=0$: the splay state and the full synchrony.
The solution corresponding with $H_D\neq0$ corresponds to SCPS.
We now discuss in detail these three cases.

\subsection{Stability of the splay state}

The splay state is characterized by $P_D(\theta,t)=1/(2\pi)$ and $\Omega_D=0$.
Let us consider an infinitesimal perturbation $u(\theta,t)$ of $P_D$.
The corresponding linearized equation is
\begin{equation}
 \frac{\partial u}{\partial t}=-\frac{1}{2\pi}\frac{\partial }{\partial \theta}\int\dd\psi G(\psi-\theta)u(\psi,t)+
D\frac{\partial^2 u}{\partial \theta^2}
\end{equation}
where we have used that $\B{P}=0$.
Using the Fourier expansion
\begin{equation*}
 u(\theta,t)=\frac{1}{2\pi} \sum_{k=-\infty}^{+\infty} \tilde u(k,t)\ee^{-i\theta k}
\end{equation*}
where
\begin{equation*}
 \tilde u(k,t)=\int_0^{2\pi}u(\theta,t)\ee^{ik\theta}\dd\theta,
\end{equation*}
and solving the integral term as in Eq.~(\ref{eq:simply}),
the time evolution of $u$ reads
\begin{align*}
 &\frac{\partial u}{\partial t}(\theta,t)=\frac{1}{2\pi}\frac{\partial }{\partial t}  \sum_{k=-\infty}^{+\infty} \tilde u(k,t)\ee^{-i\theta k}\\
 &=\frac{1}{4\pi}\left(\tilde u(1,t) \ee^{-i(\theta-\gamma_1)}+2a \tilde u(2,t) \ee^{-i(2\theta-\gamma_2)} +\cc\right)\\
 &\qquad-\frac{D}{2\pi} \sum_{k=-\infty}^{+\infty} k^2 \tilde u(k,t)\ee^{-i\theta k}.
\end{align*}
The evolution equations of the Fourier modes $\tilde u(k,t)$ are then given by
\begin{align*}
 \frac{\partial \tilde u(1,t)}{\partial t}&=\left(\frac{1}{2}\ee^{i\gamma_1}-D\right)\tilde u(1,t),\\
 \frac{\partial \tilde u(2,t)}{\partial t}&=\left(a\ee^{i\gamma_2}-4D\right)\tilde u(2,t), \\
 \frac{\partial \tilde u(k,t)}{\partial t}&=-Dk^2\tilde u(k,t)	\qquad\text{if }\; |k|>2
\end{align*}
complemented by the complex conjugate equations for the negative $k$-modes.
Manifestly, the equations are diagonal. Since the eigenvalues corresponding to the eigenfunctions 
with $|k|>2$ are negative real numbers, the stability is determined only by the eigenvalues corresponding to the 
two first Fourier modes,
$\lambda_1=\frac{1}{2}\ee^{i\gamma_1}-D$ and $\lambda_2=a\ee^{i\gamma_2}-4D$.
In the particular case $\gamma_2=\pi$, the second eigenvalue has $\Real(\lambda_2)<0$
so that the stability of the splay state is controlled only by the first mode,
\begin{equation*}
 \Real(\lambda_1)=\frac{1}{2}\cos \gamma_1 -D<0.
\end{equation*}
The critical curve where $\Real(\lambda_1)=0$ corresponds to the right almost vertical line reported in 
Fig.~\ref{fig:phasediag}.

\subsection{The synchronous state and self-consistent partial synchrony}

\begin{figure}
\includegraphics[width=0.5\textwidth,clip=true]{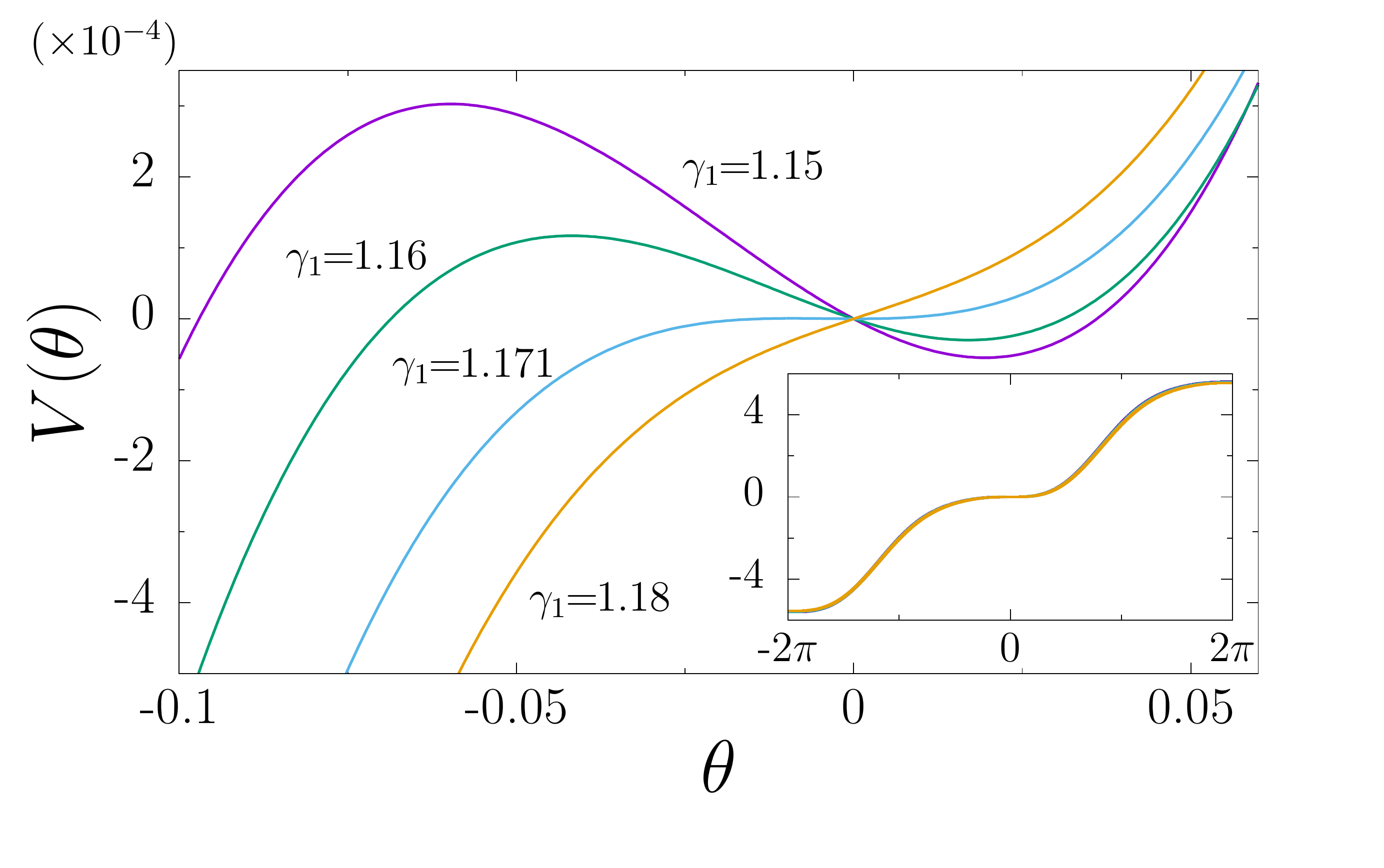}
\caption{
Washboard potentials corresponding to Eq.~(\ref{eq:potentials}) computed for a noise level of $D=6\cdot10^{-5}$.
From top left to bottom left
purple, green, blue and amber correspond to $\gamma_1=$1.15, 1.16, 1.171 and 1.18 respectively.
Inset shows the same results in a enlarged parameter range.
The parameters used in each case have been computed numerically from direct simulations with $N=10^4$ oscillators.
}
\label{fig:potentials}
\end{figure}

In the deterministic case, the fully synchronous state is characterized by a $\delta$-like distribution
and there is a well defined stability boundary for this solution.
In order to understand what happens once noise is added, it is convenient to look at Eq.~(\ref{eq:fokker-planck})
as if the velocity field were given a priori. It corresponds to a standard Fokker-Planck equation
in a washboard potential defined as
\begin{equation*}
 \frac{\partial V(\theta)}{\partial \theta}=\Omega_D-\Bt{P_D} \; ,
\end{equation*}
so that
\begin{equation}\label{eq:potentials}
 V(\theta)=
 \Omega_D\theta-\frac{1}{2}\tilde P_D(1) \ee^{-i(\theta-\gamma_1)}\!-\frac{a}{4} \tilde P_D(2) \ee^{-i(2\theta-\gamma_2)}\! + \cc
 \end{equation}
apart from an arbitrary additional constant.
Because of the tilting, as soon as $D>0$, any stationary solution is characterized by a non-zero flux $H_D$, even
if the potential has well defined minima. In other words, the fully synchronous state becomes formally equivalent
to SCPS and no transition can be any longer found. 
Nevertheless, one can still identify a sort of critical line separating the regime where the potential
has local minima and the flux is thereby driven by the noise, from the one where no minima exist and the
flux is the result of a deterministic current.
In figure~\ref{fig:potentials} we show how the potential $V(\theta)$ changes shape upon varying $\gamma_1$ 
for a fixed noise strength (the {\it unknown} parameters $\Omega_D$, $\tilde P_D(1,t)$ and $\tilde P_D(2,t)$
have been determined with the help of direct simulations).
Using this technique we have reconstructed the ``transition'' line reported in Fig.~\ref{fig:phasediag} 
(see the left quasi-vertical line). 

%%%%%%%%%%%%%%%%%%%%%%%%%%%%%%%%%%%%%%%%%%%%%%%%%%%%%%%%%%%%%%%%%%%%%%%%%%
%%%%%%%%%%%%%%%%%%%%%%%%%%%%%%%%%%%%%%%%%%%%%%%%%%%%%%%%%%%%%%%%%%%%%%%%

Having understood that, once noise is added, SCPS and full synchrony are one and the same regime, 
now we focus on the procedure to determine the shape of the stationary distribution. 
For $D=0$, i.e. in the deterministic case discussed in Ref.~\cite{Clusella-Politi-Rosenblum2016}, the
probability density can be determined from Eq.~(\ref{eq:stationary-equation}) without the need
of performing any integration,
\begin{equation}\label{eq:deterministic}
 P_0(\theta)=\frac{-H_0}{\Omega_0-\B{P_0}} \; .
\end{equation}
This solution is valid if the denominator has no zeros, i.e. if there are no minima in the corresponding potential.
The above expression depends on two complex variables $\tilde P_0(1)$, $\tilde P_0(2)$ and 
two scalars $\Omega_0$, $H_0$. Since we are free to choose the phase of the distribution $P_0$, we can assume
that $\tilde P_0(1)$ is real.  Moreover, $H_0$ can be obtained by imposing the normalization condition.
Therefore, the determination of $P_0$ requires finding a fixed point in a four-dimensional space: this problem was
tackled and solved numerically in Ref.~\cite{Clusella-Politi-Rosenblum2016}.

%%%%%%%%%%%%%%%%%%%%%%%%%%%%%%%%%%%%%%%%%%%%%%%%%%%%%%%%%%%%%%%%%%%%%%%%%%%%%%%%%%%%%%%%%%%%%%%%%%%%%%%%%%%%%%%%
In the general case $D>0$, the stationary solution $P_D$ must be obtained by integrating the
ODE~(\ref{eq:stationary-equation}). One could obtain an explicit expression for $P_D$ 
by introducing a Jacobi-Anger expansion. Such expression would involve series with terms depending 
on Bessel functions, making both the analytic and numerical treatment highly complex
and, thus, inappropriate to effectively study $P_D$ and its stability properties.

\begin{figure}
\includegraphics[width=0.45\textwidth,clip=true]{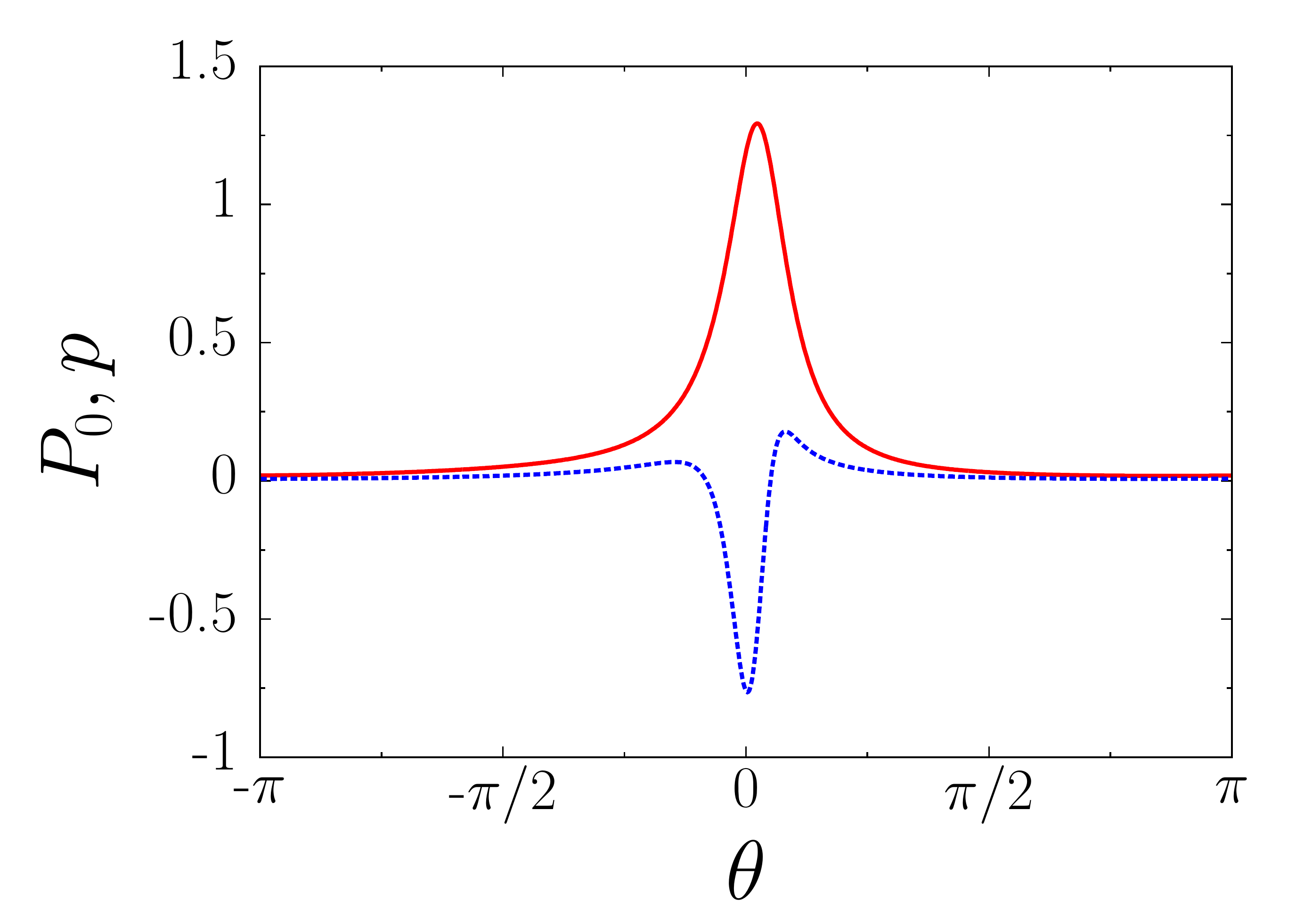}
\caption{Shape of $P_0(\theta)$ (red line) and $p(\theta)$ (blue dashed line)
for $\gamma_1=1.33$. The noise-induced perturbation $p$ has been arbitrarily rescaled.
These functions correspond to the solutions of Eq.~(\ref{eq:deterministic}) and Eq.~(\ref{eq:perturbation}) 
respectively.
}
\label{fig:solutions}
\end{figure}

It is more convenient to develop a perturbative formalism to investigate the problem in a semi-analytic
way in the limit of small noise, i.e. $D\ll 1$. At first order, the probability density can be written as
$P_D(\phi)=P_0(\phi)+Dp(\phi)$. Moreover, since we expect variations of the macroscopic frequency as well as of
the flux, we assume $\Omega_D=\Omega_0+D\omega$ and $H_D=H_0+D\eta$.
Upon replacing these assumptions into Eq.~(\ref{eq:stationary-equation}) and retaining 
first order corrections in $D$, it is found that
\begin{equation*} 
\eta=\frac{H_0}{P_0(\theta)} p(\theta)- P_0'(\theta) - \Bigl\{\omega - \B{p}\Bigr\}P_0(\theta)
\end{equation*}
where $\tilde p(k)$ stands for the $k$-th Fourier mode of $p$, while the prime denotes a derivative 
with respect to $\theta$.
This equation can be easily solved for $p(\theta)$, obtaining 
\begin{equation}\label{eq:perturbation}
 p(\theta)\!=\!\frac{P_0(\theta)}{H_0} \Biggl (\eta + P_0'(\theta)+ \Bigl\{\omega - 
\B{p}\Bigr\} P_0(\theta) \Biggr) \; . 
\end{equation}
In order to complete the identification of the solution, it is necessary to determine
$\tilde p(1), \tilde p(2)$ and the two scalars $\omega$, $\eta$.
Since $\tilde P(1)$ in equation~(\ref{eq:deterministic}) is real, so has to be $\tilde p(1)$, while
$\eta$ can be obtained by imposing the ``normalization" condition
$\int p(\psi)\dd \psi =0$. 
Therefore we are facing a problem of the same complexity as in Eq.~(\ref{eq:deterministic});
it can be solved using similar procedures.
An example of $P_0$ and of the corresponding variation $p$ obtained by solving 
Eqs.~(\ref{eq:deterministic},\ref{eq:perturbation}), is shown in Fig.~\ref{fig:solutions}. 
From the shape of $p$, we see that
the effect of noise is to deplete the left shoulder of the density and to raise a bit its tails.

%%%%%%%%%%%%%%%%%%%%%%%%%%%%%%%%%%%%%%%%%%%%%%%%%%%%%%%%%%%%%%%%%%%%%%%%%
%%%%%%%%%%%%%%%%%%%%%%%%%%%%%%%%%%%%%%%%%%%%%%%%%%%%%%%%%%%%%%%%%%%%%%%%%%

\subsection{Stability analysis of SCPS}
An accurate estimate of $P_D$ is a necessary requisite for a reliable stability analysis. In fact, only the 
stationary solution is marginally stable against a rigid translation.

Upon linearizing Eq.~(\ref{eq:fokker-planck}) around $P_D(\theta)$, we find that an infinitesimal perturbation
$u(\theta,t)$ satisfies the equation
\begin{align*}
 \frac{\partial}{\partial t}u(\theta,t)&=\frac{\partial}{\partial \theta}\Biggl(\left\{  
\Omega_D-\B{P_D}\right\}u(\theta,t)\\
 &-P_D(\theta) \Bt{u} \Biggr)+D\frac{\partial^2}{\partial \theta^2}u(\theta,t).
\end{align*}
The spectrum of this linear operator is purely point-like and can be determined by approximating 
the infinite-dimensional operator with finite matrices of increasing
size. The most effective method consists in expanding $u(\theta,t)$ into Fourier modes
\[
u(\theta,t)=\frac{1}{2\pi}\sum_{k=-\infty}^\infty \tilde u(k,t)\ee^{-ik\theta} \, .
\]
By making use of Eq.~(\ref{eq:simply}), we can rewrite the evolution equation as
\begin{equation}\label{eq:linear-stability}
\dot{ \mathbf{\tilde{\mathbf u}}} = \bigl[ \mathcal{S} (\mathbf{\tilde P_D},\Omega) - 
D\mathcal{C}\bigr]\mathbf{\tilde u}  
\end{equation}
where $\mathbf{\tilde u}=\{\tilde u(k,t)\}_{k=-\infty}^{\infty}$, 
$\mathbf{\tilde P_D}=\{\tilde P_D(k)\}_{k=-\infty}^{\infty}$,
while $\mathcal{S}$ and $\mathcal{C}$ are two matrices defined as follows
\begin{align*}
 \mathcal{S}_{mn}&(\mathbf{\tilde P_D},\Omega_D):=m\Biggl\{-\ii\Omega_D \delta_{m,n} \nonumber \\
& + \frac{1}{2} \ee^{\ii\gamma_1}[ \tilde P_D(1) \delta_{m-1,n}+ \tilde P_D(m-1)\delta_{1,n}]  \nonumber\\
 &-\frac{1}{2} \ee^{-\ii\gamma_1}[ \tilde P_D(-1)\delta_{m+1,n}+ \tilde P_D(m+1)\delta_{-1,n}] \nonumber\\
 &+\frac{1}{2} a\ee^{\ii\gamma_2}[ \tilde P_D(2) \delta_{m-2,n}+ \tilde P_D(m-2)\delta_{2,n}] \nonumber\\
&-\frac{1}{2} a\ee^{-\ii\gamma_2}[ \tilde P_D(-2)\delta_{m+2,n}+\tilde P_D(m+2)\delta_{-2,n}]  \Biggr\},
\end{align*}
\begin{equation*}
\mathcal{C}_{mn} = m^2 \delta_{m,n}
\end{equation*}

By recalling that $\tilde P_D(m)=\tilde P_0(m)+D\tilde p(m)$,
we can insert this perturbative expression into Eq.~(\ref{eq:linear-stability}), obtaining
\begin{equation}\label{eq:linear2}
\dot{\mathbf{\tilde{\mathbf u}}}  = \mathcal{S} (\mathbf{\tilde P_0},\Omega_0)\mathbf{\tilde u} +
D(\mathcal{S}(\mathbf{\tilde p},\omega)- \mathcal{C})\mathbf{\tilde u}  \; .
\end{equation}
The stability of SCPS is determined by the eigenvalues $\{\Lambda^{(k)}_D\}_k$ of the linear 
operator defined by this equation. 
Observe that $\dot{ \tilde{ u}}(0,t)=0$, while $\tilde u(0,t)$ is decoupled from all other Fourier modes. 
This is a consequence of norm conservation, which implies that one zero eigenvalue is always present. 
Another obvious property of the linear operator is that exchanging positive with negative components 
is equivalent to taking the complex conjugate: this is due to the probability density being a real quantity.

In the limit $D\to 0$ the evolution operator reduces to $\mathcal{S} (\mathbf{\tilde P_0},\Omega_0)$.
Its eigenvalues $\Lambda^{(k)}_0$ are responsible for the instability of SCPS in the zero-noise limit, 
reported in the previous section. 

The perturbative corrections to the evolution operator defined in Eq.~(\ref{eq:linear-stability})
are composed of two contributions: the first one is due to the change of shape of the probability density induced by noise; 
the second one is the direct consequence of the diffusion term, which strongly damps
short-wavelength Fourier modes (see the negative diagonal terms of $\mathcal{C}$ proportional to $m^2$).
The eigenvalues $\{\Lambda^{(k)}_D\}_k$ can be determined perturbatively by following standard procedures.
Upon expanding the eigenvalues up to first order in $D$, $\Lambda^{(k)}_D=\Lambda^{(k)}_0+D\lambda^{(k)}$, 
one can indeed write~\cite{Wilkinson1965}
\begin{equation}\label{eq:perturb}
 \lambda^{(k)}=\frac{{\mathbf{W}_0^{(k)}}^T \left[\mathcal{S}(\mathbf{\tilde p},\omega)- 
\mathcal{C}\right]\mathbf{V}_0^{(k)}}{{\mathbf{W}_0^{(k)}}^T \mathbf{V}_0^{(k)}}
\end{equation}
where $\mathbf{W}_0$ and $\mathbf{V}_0$ are the left-, resp. right-hand eigenvectors associated 
to $\Lambda^{(k)}_0$.

In principle, the matrix indices range from $-\infty$ to $+\infty$.
One can however obtain sufficiently accurate estimates of the largest eigenvalue by restricting the analysis to
a finite range $[-L,L]$, provided that $L$ is large enough. We have verified that $L=100$ suffices to reconstruct 
the relevant part of the stability spectrum.
In practice, one first needs to solve numerically the Eqs.~(\ref{eq:deterministic},\ref{eq:perturbation}) to determine 
the stationary solution of the Fokker-Planck equation (up to first order in $D$). 
Then, left and right eigenvectors of the unperturbed evolution operator are obtained, so that
we are finally able to determine the corrections to the eigenvalues.

\begin{figure}
\includegraphics[width=0.47\textwidth,clip=true]{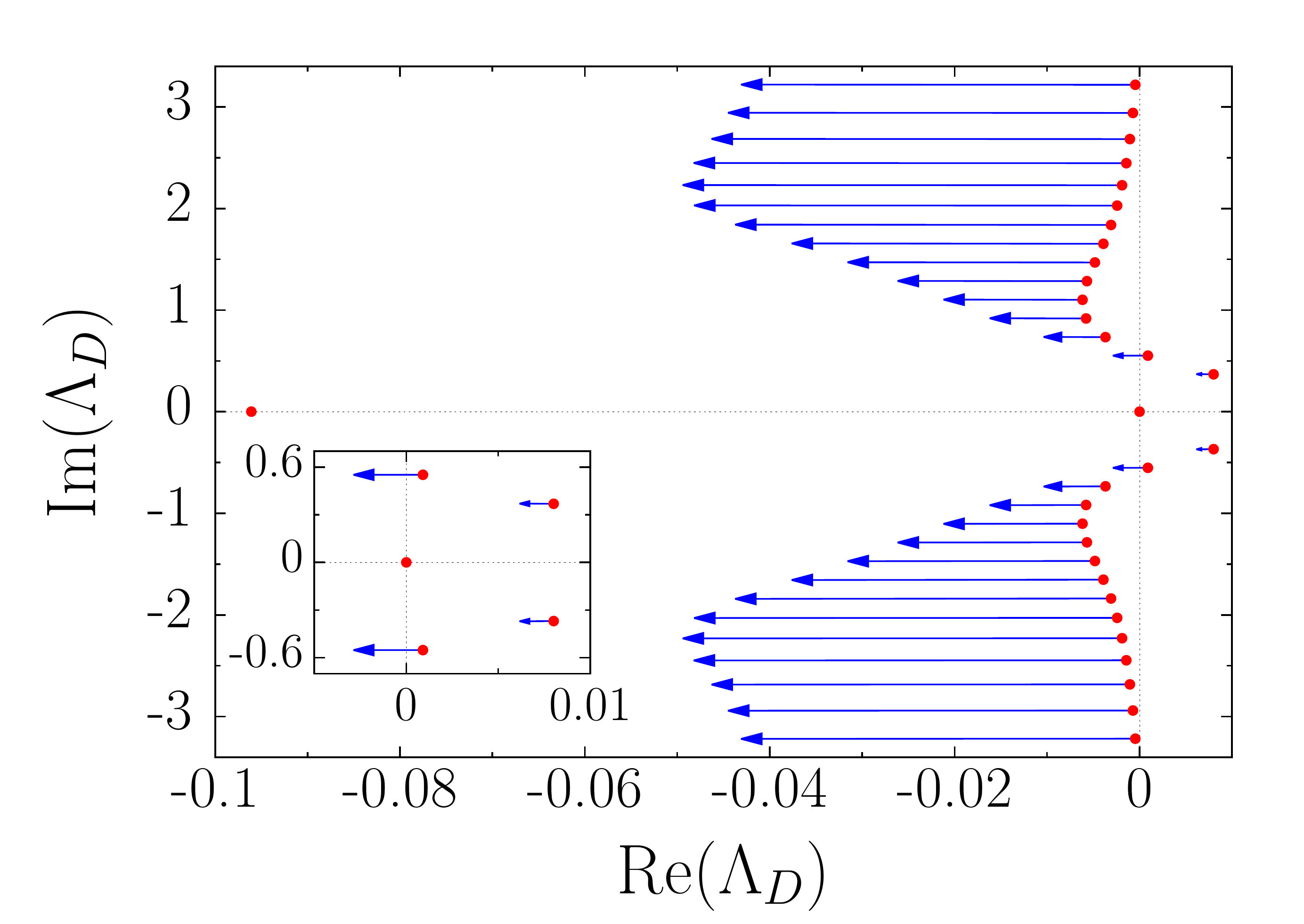}
\caption{Red circles show real and imaginary part of the eigenvalues controlling stability of $\tilde P$ for $\gamma_1=1.33$ and $D=0$.
Blue arrows show the directions where the eigenvalues are pushed when $D=1.7\cdot 10^{-5}$.
The inset is a zoom for $\Real(\Lambda_D)>0$.}
\label{fig:eigenvalues}
\end{figure}

The results are illustrated in Fig.~\ref{fig:eigenvalues}, where the eigenvalues are
plotted for $D=0$ (red dots) and $D=1.7 \cdot 10^{-5}$ (the arrows denote the shift of each eigenvalue).
The spectrum is composed of pairs of complex conjugate eigenvalues; two of them are characterized by a strictly zero
imaginary part: the first one corresponds to the zero exponent, which follows from the invariance of the solution under
a rigid phase shift and is present in the noisy regime as well; the second one is the most negative eigenvalue (see the
leftmost red dot) - very weakly affected by noise.

In the deterministic case the imaginary parts of the eigenvalues are almost equispaced.
They can be used to parametrize the eigenvectors, as $\mathrm{Im}(\Lambda_D)$ is proportional to the wavenumber $k$
(see Ref.~\cite{Clusella-Politi-Rosenblum2016}). Moreover, for $D=0$ the real parts decrease 
exponentially for increasing $k$~\cite{Clusella-Politi-Rosenblum2016}.
This is a crucial difference with mono-harmonic systems such as the Kuramoto-Sakaguchi model, where 
the conservation laws enforced by the Watanabe-Strogatz theorem~\cite{Watanabe-Strogatz-93,*Watanabe-Strogatz-94}
imply the existence of infinitely many strictly-imaginary eigenvalues.

The effect of noise is to basically shift all eigenvalues towards more negative values:
in fact, all of the arrows reported in Fig.~\ref{fig:eigenvalues} are practically horizontal and point to the left.
The stabilizing effect depends strongly on $k$ (because of the diffusive term in Eq.~\ref{eq:fokker-planck});
it is approximately proportional to $k^2$. In particular, it is very weak for the two pairs of unstable eigenvalues 
(see also the inset). Nevertheless, even for a maximally unstable $\gamma_1$  (the value chosen in
Fig.~\ref{fig:eigenvalues}), a tiny amount of noise is sufficient to stabilize one of the two pairs of unstable directions
(a noise amplitude about 7 times larger would fully stabilize SCPS).

%%%%%%%%%%%%%%%%%%%%%%%%%%%%%%%%%%%%%%%%%%%%%%%%%%%%%%%%%%%%%%%%%%%%%%%%%
%%%%%%%%%%%%%%%%%%%%%%%%%%%%%%%%%%%%%%%%%%%%%%%%%%%%%%%%%%%%%%%%%%%%%%%%%%

\begin{figure}
\includegraphics[width=0.48\textwidth,clip=true]{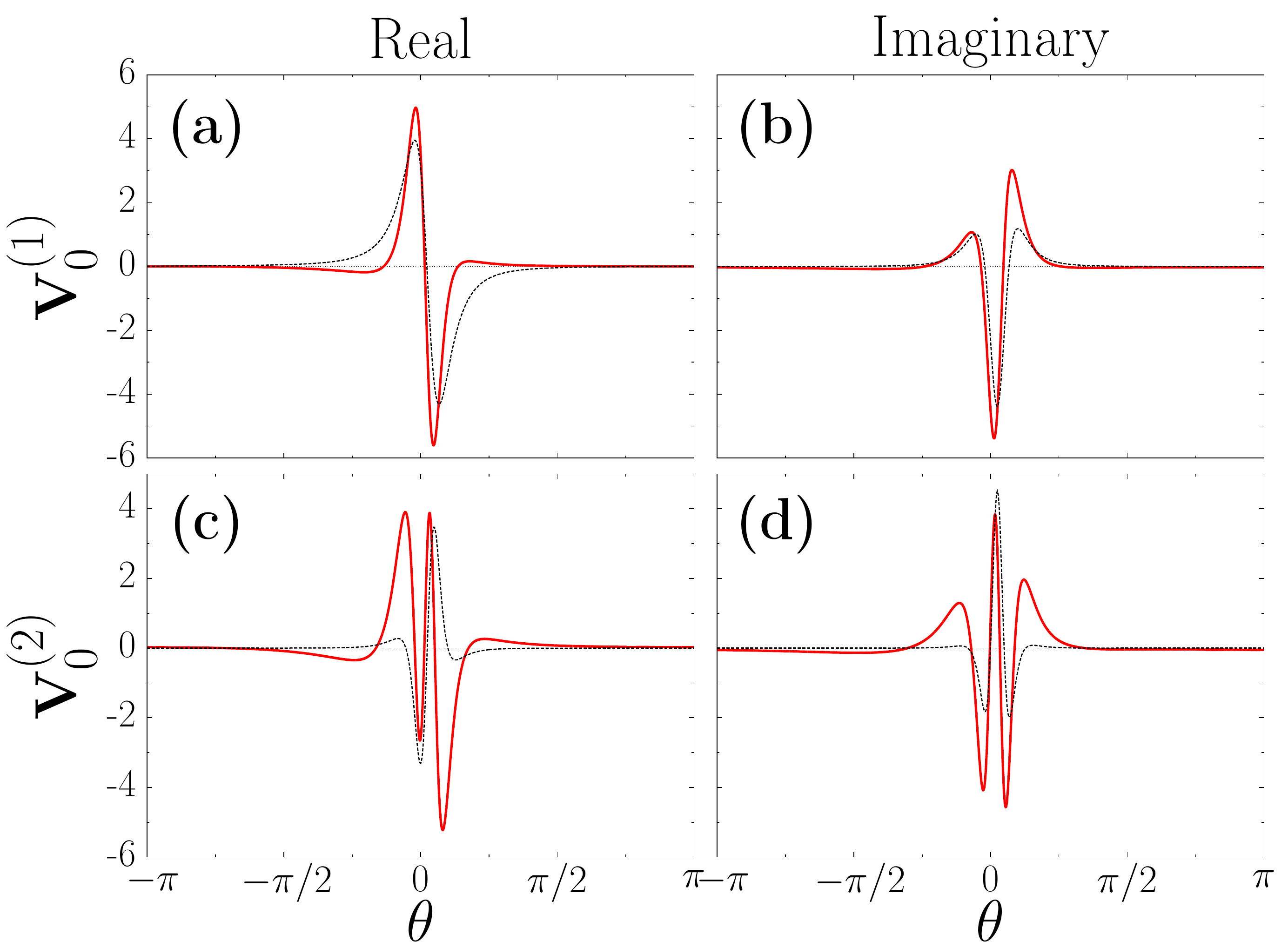}
\caption{ Eigenfunctions associated to the eigenvalues controlling the stability of SCPS for $\gamma=1.33$ and $D=0$.
Panels (a) and (b) show real and imaginary part of the eigenfunction associated to $\Lambda_0^{(1)}$ (red lines),
and first and second derivatives of $P_0$ suitable rescaled (black dashed lines) respectively.
Panels (c) and (d) show real and imaginary part of the eigenfunction associated to $\Lambda_0^{(2)}$ (red lines),
and third and forth derivatives of $P_0$ suitable rescaled (black dashed lines), respectively.
}
\label{fig:eigenfunctions}
\end{figure}

Additional information can be obtained by looking at the eigenvectors. Unsurprisingly they are 
localized in the region where the probability density is concentrated. Moreover, the higher the imaginary part
of an eigenvalue, the larger the number of oscillations of the corresponding eigenvector: this is 
a manifestation of the above mentioned (approximate) relationship beteen imaginary parts and wavenumbers.
The eigenvectors corresponding to the unstable directions,
$\mathbf{V}_0^{(1)}$ and $\mathbf{V}_0^{(2)}$ are plotted in Fig.~\ref{fig:eigenfunctions} for $D=0$, separating the 
real from the imaginary component (see the solid lines). There we also see that $\mathbf{V}_0^{(1)}$ is also reminiscent
of the first and second derivatives of $P_0$ respectively (compare red continuous and 
black dashed lines in panels (a) and (b) in Fig.~\ref{fig:eigenfunctions}).
The analogy extends to $\mathbf{V}_0^{(2)}$, if compared with the third and fourth derivatives of $P_0$, although
it is much more qualitative (see panels (c) and (d) in Fig.~\ref{fig:eigenfunctions}).

%%%%%%%%%%%%%%%%%%%%%%%%%%%%%%%%%%%%%%%%%%%%%%%%%%%%%%%%%%%%%%%%%%%%%%%%%
%%%%%%%%%%%%%%%%%%%%%%%%%%%%%%%%%%%%%%%%%%%%%%%%%%%%%%%%%%%%%%%%%%%%%%%%%%

Upon tracking the eigenvalues with positive real part for increasing $D$ we can identify
where the stabilization of SCPS takes place. The transition points are plotted as blue triangles in figure~\ref{fig:phasediag}
and show a good agreement with the direct numerical simulations.

\section{Discussion and conclusions}\label{section:4}

In the previous section we have performed a detailed stability analysis of SCPS in the presence of noise. 
This collective regime corresponds to a stationary state in a suitably moving frame, where its evolution is described 
by a nonlinear Fokker-Planck equation. The nonlinearities play a double role: on the one hand they contribute
to a self-consistent determination of the underlying potential; on the other hand they contribute to the stability
of the state itself. In fact, the dynamics is not purely drift-diffusion driven; as shown in~\cite{Clusella-Politi-Rosenblum2016} there
may be unstable directions. In the previous section we have, however, seen that even a very small noise is sufficient
to stabilize the collective dynamics.

\begin{figure}
\includegraphics[width=0.5\textwidth,clip=true]{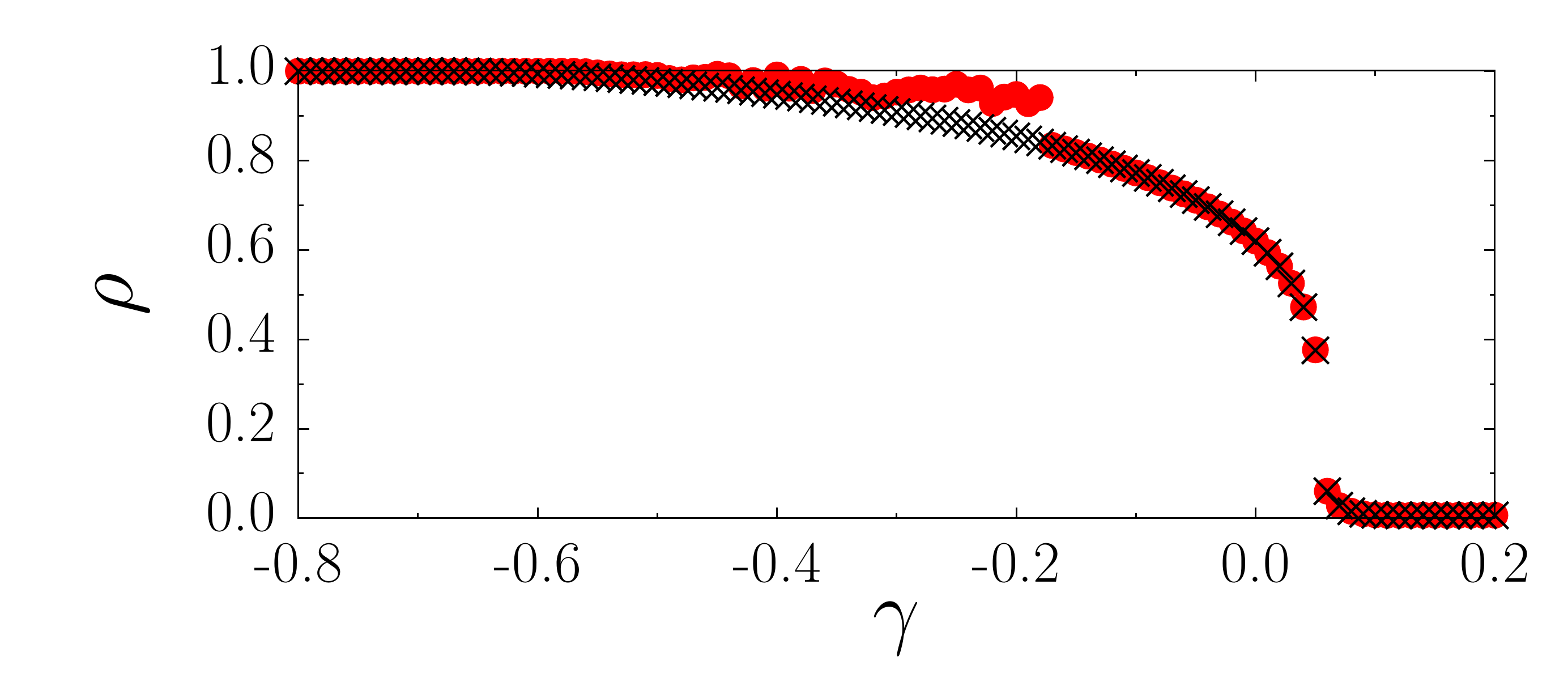}
\caption{Order parameter $\rho$ for different values of $\gamma_1$ for no noise (red circles)
and $D=1\cdot 10^{-6}$ (black crosses).
The results have been obtained by direct simulations of the Rayleigh oscillators with 1000 units.}
\label{fig:rayleigh}
\end{figure}

It is therefore natural to ask whether this scenario is peculiar of the biharmonic setup.
We check this point by studying another model, an ensemble of mean-field coupled Rayleigh oscillators, where
SCPS has been observed and found to lose stability in a purely deterministic setup~\cite{Clusella-Politi-Rosenblum2016}.
We show that a small amount of noise is again able to stabilize SCPS.
The Rayleigh oscillator model reads as
\begin{equation}\label{eq:rayleigh}
 \ddot x_j-\zeta (1-\dot x_j^2)+x_j=\varepsilon \Real[\ee^{i\gamma}(X+i Y)]+\eta_j(t)
\end{equation}
where $X=N^{-1}\sum_m x_m$ and $Y=N^{-1}\sum_m\dot x_m$ are the mean field contributions to the coupling, and 
$\varepsilon=0.05$ is the coupling strength.
We assume again white noise $\eta_j(t)$  with $\langle \eta_j(t) \rangle = 0$ and 
$\langle \eta_j(t)\eta_m(t')\rangle=2D\delta_{jm}\delta(t-t')$.
The parameter $\zeta$ determines the stability of the limit-cycle.
In this work we discuss the case $\zeta=5$ for which there is a strong attraction.
Therefore $\gamma$ is the main control parameter that is going to be tuned. 
An appropriate order parameter is
\begin{equation*}
 \rho=\text{rms}(X)/\text{rms}(x)
\end{equation*}
where
\begin{equation*}
 \text{rms}(x)=\sqrt{\langle x(t)^2 \rangle} 
\end{equation*}
is the root mean square of the time evolution.
Therefore $\rho=1$ when there is fully synchrony and $\rho=0$ when the oscillators are distributed uniformly.
In the deterministic case, in the range $\gamma\in[-0.7, 0.2]$ a wide number of dynamical regimes are observed 
(see red circles in figure~\ref{fig:rayleigh}). In particular,  SCPS is observed for $-0.18\lesssim \gamma \lesssim 0.05$.
Above $\gamma\simeq 0.05$ the system converges to a homogeneous nine-cluster state.
Similarly to the biharmonic model, SCPS loses stability towards a two-cluster state for $\gamma\simeq -0.18$.
This cluster state finally converges to full synchrony at $\gamma\simeq -0.57$.
On the one hand, a small noise ($D=1\cdot10^{-6}$) does not substantially affect the regions where full synchrony and the nine-cluster 
states are stable.  On the other hand, it is once again able to stabilize SCPS in the entire 
interval up to full synchrony (see the black crosses in Fig.~\ref{fig:rayleigh}).

%%%%%%%%%%%%%%%%%%%%%%%%%%%%%%%%%%%%%%%%%%%%%%%%%%%%%%%%%%%%%%%%%%%%%%%%%
%%%%%%%%%%%%%%%%%%%%%%%%%%%%%%%%%%%%%%%%%%%%%%%%%%%%%%%%%%%%%%%%%%%%%%%%%%

The study of two models of phase oscillators has shown that a small amount of microscopic noise 
stabilizes self-consistent partial synchrony. It is natural to ask whether this effect extends to
other types of collective dynamics.
In globally coupled identical maps, collective chaos can be observed~\cite{Shibata-Chawanya-Kaneko-1999}.
In such a setup, it was found that an additive noise of the type considered in this paper 
can reduce the dimensionality of the collective dynamics~\cite{Shibata-Chawanya-Kaneko-1999,Takeuchi-Chate2013}. 
Considering that a chaotic evolution can
be seen as a sort of wandering process among different unstable periodic orbits, it is tempting to
interpret this reduction of dimensionality as a progressive stabilization of the dynamics along various
directions. It will be instructive to further investigate this interpretation.

\section*{Acknowledgement}
This work has been financially supported by the EU project COSMOS (642563).

\bibliography{references_noise}

\end{document}